\documentclass[letterpaper,10pt]{article}

\usepackage{amsmath}
\usepackage{amssymb}

\usepackage{graphicx}
\usepackage{cite}

\usepackage{color} 
\usepackage[small]{caption}

\makeatletter
\renewcommand{\@biblabel}[1]{\quad#1.}
\makeatother

\date{}

\begin{document}

\begin{flushleft}
{\Large
\textbf{Quantifying loopy network architectures}
}
\\
Eleni Katifori$^{1,\ast}$, 
Marcelo Magnasco$^{1}$, 
\\
\bf{1} Laboratory of Mathematical Physics, The Rockefeller University, New York, NY, USA
\\

$\ast$ E-mail: ekatifori@rockefeller.edu
\end{flushleft}

\begin{abstract} 
Biology presents many examples of planar distribution and structural networks having dense sets of closed loops. An archetype of this form of network organization is the vasculature of dicotyledonous leaves, 
which showcases a hierarchically-nested architecture containing closed loops at many different levels. 
Although a number of methods have been proposed to measure aspects of the structure of such networks, a robust metric to quantify their hierarchical organization is still lacking. 
We present an algorithmic framework,  the {\sl hierarchical loop decomposition}, that allows mapping loopy networks to binary trees, preserving in the connectivity of the trees the architecture of the original graph. 
We apply this framework to investigate computer generated graphs, such as artificial models and optimal distribution networks, as well as natural graphs extracted from digitized images of dicotyledonous leaves and vasculature of rat cerebral neocortex. We calculate various metrics based on the Asymmetry, the cumulative size distribution and the Strahler bifurcation ratios of the corresponding trees and discuss the relationship of these quantities to the architectural organization of the original graphs. This algorithmic framework decouples the geometric information (exact location of edges and nodes) from the metric topology (connectivity and edge weight) and it ultimately allows us to perform a quantitative statistical comparison between predictions of theoretical models and naturally occurring loopy graphs. 
\end{abstract}

\section{Introduction}
Among the many different classes of complex systems that can primarily be described as ``networks'', an important subclass concerns physical networks devoted to transportation of various entities, such as fluids or energy. To some extent, structural load-bearing networks can also be considered in this category, as their job is the distribution of stress-strain. Besides their evident technological importance, these networks are central to the function of living beings; because of their concrete physicality they are sometimes far more accessible to experimental analysis than other important biological networks, and hence offer an important window into the organization and function of naturally evolved large-scale networks.

\begin{figure}[b!] 
\begin{center}
\includegraphics[width=3.4in]{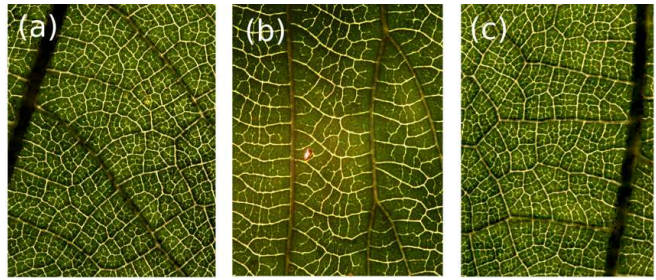}
\caption{\label{figure1}
\textbf{Variability in natural loopy networks.} (a), (b) Leaf vasculature of two dicotyledonous species. (c) Detail of leaf collected from the same plant as leaf (a). The venation of (a) and (c) is predominately reticulate, (b) is percurrent. In general, leaves from the same plant (or species) share statistically similar architectural properties, as compared to leaves from different species. The scale is 1 cm.
} 
\end{center}
\end{figure}

Many biological distribution and structural networks contain dense numbers of reentrant loops. The venation of angiosperm leaves (Fig.~\ref{figure1}) \cite{Ellis2009}, the structural veins of insect wings, the continuously adapting foraging networks of some fungi and slime molds \cite{Tero2010},  the vasculature of  animal organs such as the adrenal glands, the brain \cite{Blinder:2010p2746} and the liver are just a few of a large number of examples where physical networks developed loops in living organisms. These networks perform functions crucial to the survival of the organisms that use them. The hierarchical organization and the intricacies of the architecture of these highly interconnected networks dictate the efficacy in providing support or distributing load under varying conditions. In some cases the function of closed loops and how many there should be is intuitively obvious; the webbing-like veins of a dragonfly wing have cross-bracings that serve to maintain rigidity and resistance with a minimum of weight. In other cases it is not self-evident why there are as many loops as observed.

In many cases, such as leaf venation, loopy networks evolved gradually from a tree architecture \cite{Melville1969}. Various reasons for the evolution of loopiness in biological distribution networks have been proposed \cite{Katifori:2010p2741, Corson:2010p2818, Roth-Nebelsick2001a}. 
These networks are the result of developmental processes that frequently dictate not the exact position of each network edge but the overall organization in a statistical sense. 
For example, one can frequently determine by mere inspection of the leaf venation patterns if two leaves are specimens from two different species (Fig.~\ref{figure1}). Similarly, networks produced {\sl in silico} by optimization routines or developmental simulations that incorporate the effects of biological noise exhibit architectures that are to some extent random: each simulation repeat will produce statistically similar, but never identical, networks \cite{ Couder2002, Dimitrov2006, Fujita2006,  Rinaldo2006, Corson:2009p3346}. To compare naturally occurring networks with the computer simulated models we therefore need to be able to test the null hypothesis that the two networks in question have been drawn from the same distribution. 


Some of the distribution and structural networks in question are planar, i.e. their edges are (or can be) all confined to the same plane and meet only at vertices (no two edges can cross each other). Examples of naturally occurring planar networks include the veins of leaves and insect wings, the loopy arterial network of the mammalian neocortex and many others.

Despite the importance of  these planar loopy networks, the arsenal of specialized tools and techniques that can sufficiently capture the architecture is still limited. Instead, so far the scientific focus has been on quantifying and describing the topology of networks with a tree architecture or the connectivity of non-planar complex networks, such as the internet. In particular, work developed since the fifties to describe river networks and dendritic architectures helped establish some powerful measures to describe the topological properties of tree structures. The Horton-Strahler stream ordering system \cite{HORTON:1945p3487, STRAHLER:1952p3188} and the Asymmetry \cite{VANPELT:1992p2856} are two such measures that played a crucial role in understanding the laws that dictate network growth and organization. Invaluable though these measures might be for rivers and dendrites, their definition and usage presuposes a tree architecture and loops destroy their consistency.

Although measures developed for general, non-planar complex networks such as the degree distribution and the community structure in principle work for planar graphs, frequently they are not fine tuned to capture many aspects of the 2-d network organization \cite{Rinaldo:1998p259, Albert2002, Newman2003, Barrat2004, Boccaletti:2006p256, Costa2007, Shao:2009p3561}. Methods to extract the hierarchical organization of complex networks have focused primarily on the node connectivity \cite{Sales-Pardo2007}. Similarly, other more specialized metrics such as the distribution network entropy \cite{Ang2005} are not very informative with regard to quantities of interest in this work, and in particular the hierarchical organization of graphs. Some specialized schemes have been developed to quantify the loopy architecture of dicotyledonous leaves (see e.g. \cite{Rolland-Lagan2009}), and though they can reveal important information about leaf physiology and function \cite{Blonder:2011p3510} these methods do not explicitly characterize the nestedness of the topology.

To achieve a meaningful and elegant quantification of highly interconnected and loopy biological networks we need a sufficiently nuanced metric that captures certain important aspects of the topology and architecture of the loopy network without relying on the exact value of the bond strength or geometrical location. Such a metric would allow phenotypic parameter reduction and assignment of numeric values to the level of loop nestedness and other aspects of the architectural organization that are not represented by descriptions that rely on local, scalar quantities (such as histograms of the vein density).
More importantly, it will allow a quantitative and topologically based comparison between natural loopy networks and the prediction of optimization models.

In this paper we present a method that allows us to map the architectural organization of a planar graph to that of a binary tree. We then use three metrics widely used for binary trees and examine their properties with regard to the original graph.. These metrics are the Asymmetry \cite{VANPELT:1992p2856}, the cumulative size distribution \cite{TAKAYASU:1988p3366, Paik:2007p946} and the Strahler bifurcation ratio \cite{STRAHLER:1952p3188}. We present results from three classes of networks: computer generated networks (whose building rules are predetermined), networks optimized for known functionals and naturally occurring networks such as leaf veins and the arterial vasculature of the rat neocortex. We finally discuss the advantages and disadvantages of each approach and present future directions and applications.

\section{Results}

\subsection{Hierarchical decomposition}

We have developed a method that maps a predominately loopy architecture to a dichotomously branching tree. This method hierarchically decomposes the loopy architecture by succesively deleting edges and joining contiguous loops, and represents this hierarchical decomposition as a tree, termed the \textit{nesting tree}.

\begin{figure*}[t!] 
\begin{center}
\includegraphics[width=6.8in]{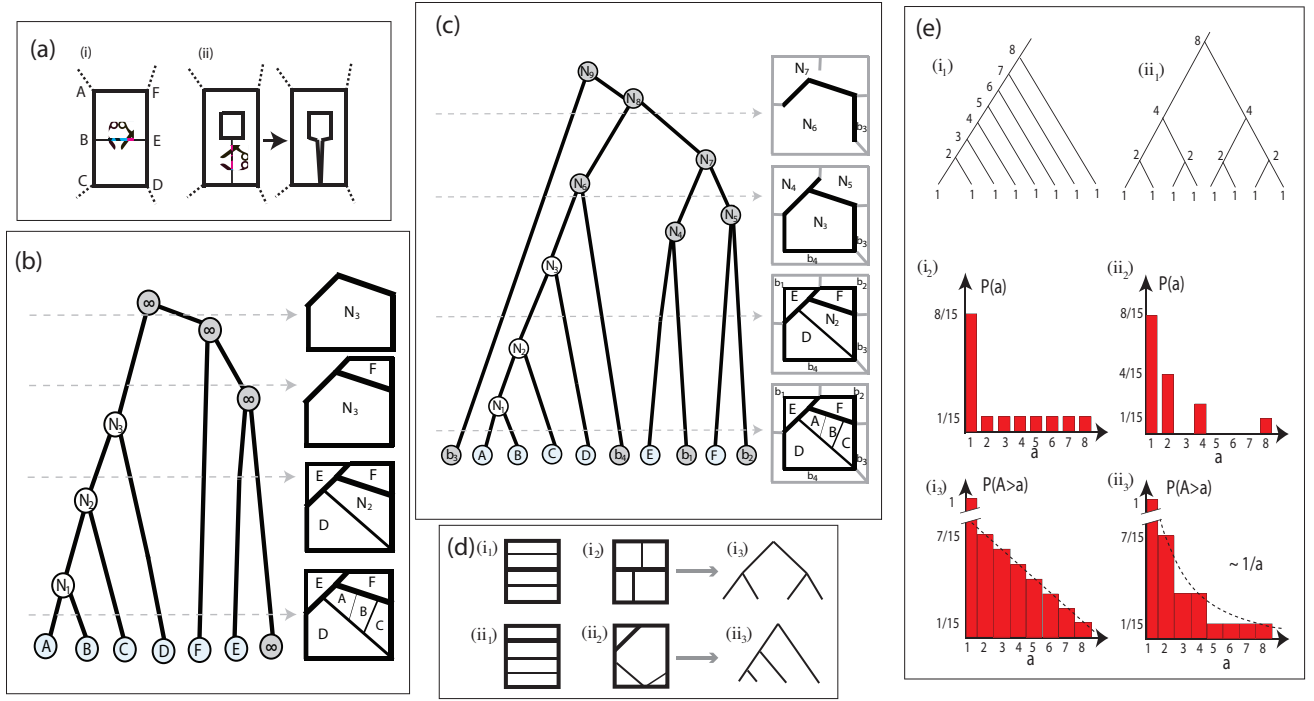}
\caption{\label{figure2}
\textbf{(a)} Deletion of an edge in a loopy graph. (i) The deletion of the edge joins two adjacent loops. (ii) The deletion of the edge disconnects the graph.
\textbf{(b)}Hierarchical decomposition of a planar graph. Boundary loops sequentially join the outside space, marked as $\infty$. Left: Nesting tree of the hierarchical decomposition. Right, top to bottom: hierarchically decomposed graph. The bottom right panel corresponds to the full graph, the rest represents the network at different levels of decomposition (the corresponding cutoff level of the tree representation is marked with a gray dashed arrow). As edges of the graph are hierarchically deleted, based on their thickness, the original loops (A-E) are joined to form derived loops ($N_1$-$N_3$). 
\textbf{(c)} Hierarchical decomposition of a planar graph. Phantom boundary loops surround the graph perimeter. Loops contiguous to the perimeter of the graph join a ring of phantom boundary loops. The decomposition proceeds as in (b), but the phantom loops $b_1$-$b_4$ appear among the loops of the original graph in the tree representation.  
\textbf{(d)} Building blocks of a loopy architecture. The two basic building blocks of the loopy architecture can be identified using the tree representation of the graph. (i${}_1$),(i${}_2$):  multiplicative nestedness. Nested loops merge hierarchically. (i${}_3$): This architecture is represented by ``tall" trees. (ii${}_1$),(ii${}_2$):  additive nestedness. Ordered loops join consecutively. (ii${}_3$): The tree representation is that of ``short'' trees. Graphs (i${}_1$) and (i${}_2$) map equivalently to (i${}_3$), similarly graphs (ii${}_1$) and (ii${}_2$) map equivalently to (ii${}_3$). 
\textbf{(e)} Cumulative size distributions of additive and multiplicative models of nestedness. (i${}_1$) Nesting tree for additive nestedness. The degree of each node is is shown. (i${}_2$) Degree (size) distribution for additive nestedness.  (i${}_3$) Cumulatize size distribution for additive nestedness. (ii${}_1$) Nesting tree ,(ii${}_2$) Degree (size) distribution and (ii${}_3$) Cumulatize size distribution for multiplicative nestedness. 
} 
\end{center}
\end{figure*}

In what follows, the term \textit{link} will refer to a graph element that connects two nodes, and the term \textit{edge} will refer to a chain of links, connecting nodes. Each node in an edge is connected to exactly two other nodes, except the nodes at the boundaries of the edge, which can be connected to only one other node (when that edge is the ``leaf'' of a tree), or three or more other nodes. The ``edge strength'' $W_J$ is a quantity that parametrizes the weight of the edge $J$. If an edge $J$ is composed of a chain of links, then $W_J$ can be set to be the edge strength of the weakest of the chain links, the median value, or any other quantity that is of interest. 
The term \textit{loop} is used to refer to the graph cycles, and the \textit{terminal} or \textit{ultimate} loops are the cycles that do not contain other loops.

The first step of this hierarchical decomposition framework is a pruning step, where all tree-like components rooted on the loopy graph backbone, if present, are removed from the graph. This step eliminates all vertices that belong only to one edge, and produces a graph where each edge either separates or connects two loops (Fig.~\ref{figure2}(a)). 

In the second step of this hierarchical decomposition, we order the list of graph edges based on their width (if the graph edge is composed of a single link) or  their edge strength, and identify the edge with the smallest $W_J$.
In this step we assume that the edges can be ordered according to their weight in a strictly monotonic fashion, namely that $W_J\ne W_K$ for every pair of edges $J,K$. This is a requirement that  can easily be implemented by infinitesimally randomly perturbing $W_J$ or $W_K$ when $W_J=W_K$.

In the third step, we remove the edge $J_s$ with the smallest edge strength from the graph. When an edge separates two contiguous loops, as in Fig.~\ref{figure2}(a)(i),  then its removal will result in joining the two loops to form a larger one, the area of which is the sum of the areas of the two initial loops. In most cases, this step will also result in joining the remaining edges of the contiguous loops. For example, in Fig.~\ref{figure2}(a)(ii), the links AB and BC will be joined to form the edge AC.

We then repeat steps 1,2 and 3 iteratively, to sequentially remove every edge, and as a consequence, gradually join every loop, and perform what we have termed a \textit{hierarchical decomposition} of the graph. We can represent this procedure with a dichotomously branching tree, as follows. The ``leaves" of the tree are the original loops of the full graph, loops A-E in Fig.~\ref{figure2}(b), and each node downstream of the leaf nodes represents a larger loop that is formed by joining two upstream loops through the removal of an edge.
The location of the downstream nodes on the vertical axis of the branching tree represents the edge strength that was removed to join these two loops. Loops are being hierarchically combined until they break to the outer region, termed \textit{exterior} (and labeled $\infty$). The exterior is treated as a separate loop. 

This method will hierarchically decompose the original graph and will register this hierarchical decomposition as a binary, nesting tree. The branching patterns of this nesting tree contain information about important topological properties of the original graph.
The nesting tree allows us to adapt and use measures traditionally used and defined on trees, to quantify the architecture and topology of loopy graphs.

The hierarchical decomposition and the nesting tree contain no explicit information about the geometry of each edge and element of the graph, other than the fact that the two joining loops need to be adjacent. Nodes of the nesting tree thus correspond to \textit{neighborhoods} of the original graph - the nesting subtree $t_j$ rooted at node $j$ represents the architecture of the subgraph enclosed in the loop represented by node $j$.

When edges at the graph perimeter are removed and loops at the boundary merge with the exterior, the neighborhood information is lost. We can retain that information by appropriately fragmenting the exterior region. Instead of having a single exterior loop, where every boundary loop sequentially merges to, we define a multitude of exterior loops as follows. We consider an exterior phantom loop that encompasses the original graph in its entirety. We then connect the vertices on the perimeter of the original graph with the perimeter of the phantom loop as shown on Fig.~\ref{figure2}(c). Thus defines $n$ boundary phantom loops, labeled $b_1,..., b_n$, where $n$ is the number of loops in the original graph that are adjacent to the perimeter. The added phantom exterior loop and links are assigned infinite weights and will never be removed during the hierarchical decomposition. After the addition of the phantom loops to the original graph, we proceed to iteratively decompose the graph as before, and represent the decomposition with a binary nesting tree, like the one shown in Fig.~\ref{figure2}(b). In this way, the neighborhood information at the boundaries is preserved and will be reflected in the architecture of the nesting tree.

The nesting tree facilitates straighforward identification of the two basic building blocks of the organization of a planar graph. We will denote these building blocks as \textit{multiplicative} (Fig.~\ref{figure2}(i${}_1$)(i${}_2$)), and \textit{additive} (Fig.~\ref{figure2}(ii${}_1$)(ii${}_2$)). The multiplicative building blocks consist of events where the small loops are joined in an iterative, self-similar fashion. It maps to a tall binary tree, such as the one shown in Fig.~\ref{figure2}(i${}_{1,2}$). The additive building block is characterized by sequential joining of minor loops to an encompassing major loop. It maps to a short binary tree, as in Fig.~\ref{figure2}(ii${}_{1,2}$).

This mapping to a nesting tree is not a bijection. Any information about the geometric organization (shape and location) of the loops is lost. Only topological information is retained. For example, networks Fig.~\ref{figure2}(i${}_1$) and (i${}_2$) both map to Fig.~\ref{figure2}(i${}_3$), and Fig.~\ref{figure2}(ii${}_1$) and (ii${}_2$) both map to Fig.~\ref{figure2}(ii${}_3$). Elements of the architectural organization, such as loop area or aspect ratio can be retained by assigning related values to the nodes of the tree $j$ and defining quantities that reflect their distribution. For example, the cumulative size distribution is based on measurements of the loop areas $A(j)$ assigned to the nodes $j$ of the nesting tree.

When an edge connects, rather than separates, two loops, its deletion will disconnect the graph (Fig.~\ref{figure2}(a)(ii)). 
There is a number of ways to incorporate such an event to the hierarchical decomposition algorithm. In the example cases that we consider in this work, such events are rare, so for simplicity we chose to discard them in our implementation. In particular, we replaced the weight value of the disconnecting edges with the maximum edge width value of the disconnected loopy components, this way ensuring that the loop will not disconnect from the graph before it is hierarchically merged to the encompassing loop (Fig.~\ref{figure2}(a)(ii)).

The nesting tree allows the unique assignment of a number to properties of the hierarchical organization of the graph and decouples geometry from topology. In this paper we consider and adapt three measures that have been traditionally used on trees: the Asymmetry, the cumulative size distribution and the Strahler bifurcation ratio. We apply those measures to the nesting tree and consider what they mean for the organization of the original graph.

\subsection{Asymmetry}

The Asymmetry is a metric that characterizes the topological structure of a binary tree. It was first developed mainly in the context of neuronal branching patterns, such as dendritic trees and was defined as the weighted mean value of the asymmetry of its partitions. Adjusting the definition and notations of \cite{VANPELT:1992p2856}, we define the partition asymmetry of a bifurcation vertex $j$ as: 
\begin{equation}\label{Eq.q}
q(r_j,s_j) = \frac{s_j-r_j}{s_j}
\end{equation}
with $s_j\ge r_j$ and $s_j+r_j \ge 2$. The parameters $r_j$ and $s_j$ are the degrees of the two subtrees at partition $j$. The degree of a (sub)tree is defined here as the total number of the leaf nodes (terminal segments) of that (sub)tree. 
Note that Eq.~\ref{Eq.q} differs slightly from the definition in \cite{VANPELT:1992p2856}.

The Asymmetry $Q_T(t_n)$ of a subtree rooted at node $n$ can now be defined as the weighted average of the partition asymmetry  $q(r_j,s_j)$ of the nodes $j\in t_n$:
\begin{equation} \label{Eq.asymmetry}
Q_T(t_n) = \frac{1}{w(t_n)} \sum_{j=1}^{d(n)-1} w_j q(r_j,s_j)
\end{equation}
where $j$ runs over all $d(n)-1$ bifurcating vertices of the subtree ($d(n)$ is the degree of the subtree),  and $w_j$ is the weight of the partition $j$.
Finally, the normalization factor $w(t_n)$ is defined as: 
\begin{equation}
w(t_n) =  \sum_{j=1}^{d(n)-1} w_j.
\end{equation}

The averaged Asymmetry $\bar{Q}_T(\delta)$ of trees of degree $\delta$ is defined as:
\begin{equation} \label{Eq.average_asymmetry}
\bar{Q}_T(\delta)= \frac{1}{n_{\delta}} \sum_{\{t_j\}, d(tj)=\delta} Q_T(t_j)
\end{equation}
where $n_{\delta}$ is the number of nodes with degree $\delta$.
In this work, we adjust this definition to be the mean of the asymmetry for all the nodes whose degree is within a distance $\Delta/2$ from $\delta$:
\begin{equation}
\bar{Q}_T(\delta)\sim  \sum_{\{t_j\}, |d(tj)-\delta|\le \Delta/2} Q_T(t_j)
\end{equation}

Calculated on the nesting trees of the hierarchical decomposition, the Asymmetry is an metric that quantifies the nestedness of the original graph. High Asymmetry values correspond to a graph that is primarily composed from \textit{additive} building blocks, and low asymmetry values correspond to a graph that is made from \textit{multiplicative} building blocks. The actual correspondence between Asymmetry values and level of nestedness depends on the choice of weight function $w_j$. Different choices of weight functions amplify different aspects of the graph architecture, and comparisons of Asymmetry plots of different graphs should only be done when the weight function choice is consistent.

In the bulk of this work we will use a weighted averaging window that includes all nodes of the subtree, with weight $w_j=d(j)-1$. In the Supplemental material we present results acquired by considering no averaging $Q_0(t_n) \equiv q(r_n,s_n)$ and by averaging over a shallow averaging window.

\begin{figure*}[t!] 
\begin{center}
\includegraphics[width=6.8in]{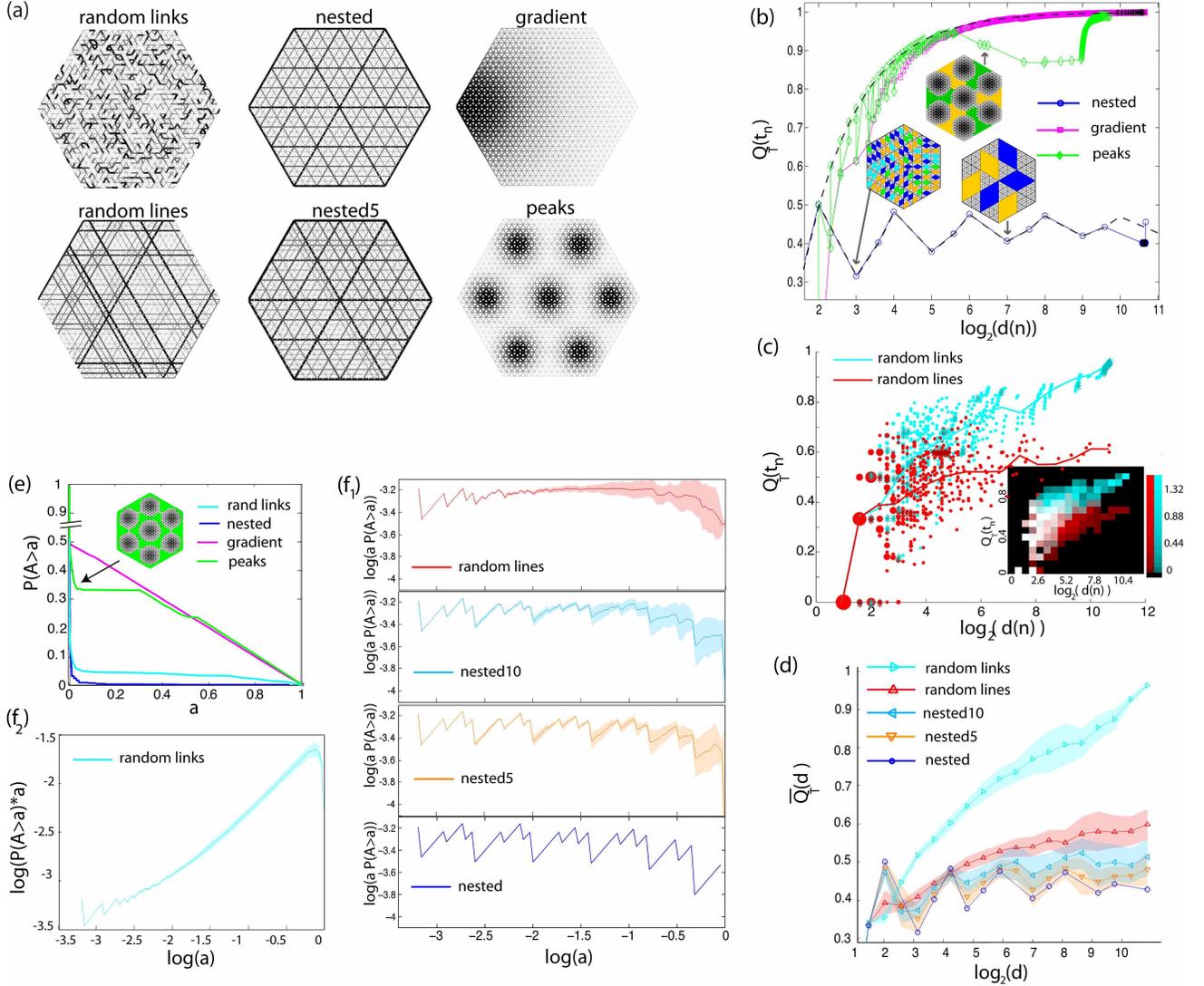}
\caption{\label{figure3}
\textbf{(a)} Generated graphs. These graphs were constructed to share identical underlying topology (N = 817 vertices, triangular lattice) and edge width distribution. 
\textbf{(b)} The Asymmetry $Q_T(t_n)$ of the every subgraph $t_n$ of rooting node $n$ is plotted as a function of the base 2 logarithm of the degree  $d(n)$, for the nested (red circles), gradient (green squares), and peaks model (blue diamonds). For the peaks and random lines model, instances of the graph are plotted with highlighted subgraphs of degree $2^3$ and $2^7$ (nested) and $\sim 2^{6.4}$ (peaks). Note the quasi-periodicity of the Asymmetry of the nested model (a signature of the self similar structure of the nested model) and the change of monotonicity of the peaks model (indicating a qualitative change in the architecture of the graph at that level of organization).
\textbf{(c)} The Asymmetry $Q_T(t_n)$ of the random lines model (red) and random links model (cyan). The x-axis is the logarithm of the degree of the vertex or the nesting tree. Red line: averaged Asymmetry of subgraphs of degree $d(n)$, random lines model. Cyan line: averaged Asymmetry of subgraphs of degree $d(n)$, random links model.
Inset: Density plots: The overlap of the two distributions is plotted in white.
\textbf{(d)} The averaged Asymmetry $\bar{Q}_T(d)$ of the nested (blue), nested5 (orange), nested10 (light blue), random lines (red) and random links model (cyan) as a function of the base 2 logarithm of the degree $d$. 
 \textbf{(e)} Cumulative size distribution $P(A>a)$ of generated models. Random links model (green), nested (blue), gradient  (magenta), peaks (green). The total area of the graphs has been normalized to 1. Discontinuities or near discontinuities in the slope of cumulative size distribution indicate lengthscales where potentially the architectural organization changes qualitatively. 
\textbf{(f${}_1$)}. Adjusted cumulative size distribution, random links model. 
\textbf{(f${}_2$)}The Adjusted cumulative size distribution $P(A>a)*a$ is plotted for the nested (blue), nested5 (orange), nested10 (light blue) and random lines model (red).The Adjusted cumulative size distribution of the self-similar networks (nested, nested5, and random lines) can be approximated by a straight line of slope zero. Notice the periodicity in the nested lines model. The colored area designated the standard error of 20 realizations.
}
 \end{center}
\end{figure*}

\clearpage

\subsection{Cumulative size distribution}

The cumulative size distribution \cite{TAKAYASU:1988p3366, Paik:2007p946} is the cumulative distribution over the areas associated with the nesting tree nodes. It is calculated by assigning an area Value $A(j)$ to each node $j$ of the nesting tree, and then calculating the probability $P(A>a)$ that an area drawn at random will exceed a certain value $a$. In general, we can associate the nesting tree nodes with any quantity that reflects a property of the original graph that is of interest, such as the total number of terminal loops nested in loop $j$ of the original graph (equal to the degree $d(j)$ of node $j$ of the nesting tree, if the terminal loops are of equal size).

The cumulative size distribution reflects the overall architecture of the original graph, as the smaller degree nodes of an aggressive subdivision, like the one in Fig.~\ref{figure2}(e)(ii) will be overepresented in the degree and cumulative degree distribution. It is easy to show that the cumulative degree distribution of iterative, self similar architectures is inversely proportional to the area 
\begin{equation}\label{Eq.self_similar}
P(A>a)\sim 1/a. 
\end{equation}
 Conversely, the cumulative degree distribution of an architecture with additive nestedness (Fig.~\ref{figure2}(e)(i)) is a straight line with slope:
\begin{equation}
\frac{dP(A>\alpha)}{d\alpha}=-\frac{1}{2}.
\end{equation}

\subsection{Strahler bifurcation ratio}

The Horton-Strahler stream-ordering system has been an invaluable tool in quantifying aspects of river topology and architecture since its inception in the fifties by Horton and Strahler \cite{HORTON:1945p3487, STRAHLER:1952p3188}. It has since been used with considerable success in describing the topology of a wide class of natural and man-made networks. 

According to the Horton-Strahler stream-ordering system, the terminal nodes of the network (the leaves) are assigned Strahler order $1$. The order of every non-leaf node is determined by the following rule: when two edges are connected to two nodes of Strahler order $\omega_1,\omega_2$ upstream, the node downstream is assigned an order
\begin{equation}
\omega = max(\omega_1,\omega_2) + \delta_{\omega_1,\omega_2}. 
\end{equation}

The Strahler numbers (or the related Horton numbers) can be used to quantify the tree topology in a number of ways. In this work we focus in particular on the Strahler bifurcation ratio, defined as:
\begin{equation}\label{Eq.Strahler_bifurcation}
 R_{\omega}=S_{\omega}/S_{\omega+1}
\end{equation}
where $S_{\omega}$ is the number of streams of Strahler order $\omega$. A stream is defined as a maximal path of branches connecting vertices of Strahler order $\omega$, ending in a vertex of higher order.

The law of stream numbers states that the stream numbers $S_{\omega}$ approximate an inverse geometric progression with the order $\omega$, a statement that implies $R_{\omega}=const$. However, it is not possible to use this law as evidence of self-similarity of a distinctive architecture, as  it is followed by the vast majority of binary trees \cite{KIRCHNER:1993p265}.

The Horton-Strahler stream-ordering system cannot be directly used to describe loopy networks, as there can be no unique assignment of the stream order in a redundant graph. The hierarchical decomposition and the nesting tree provide a mapping that allows assignment of Strahler numbers to a loopy graph, as the loops of the original graph map to the vertices of the nesting tree and the Strahler number of node $j$ depends on the nestedness of the graph segment enclosed by the loop $j$.

We now analyze examples from three classes of graphs: models generated by specific, prescribed building rules, outputs of optimization routines and natural graphs  (in particular the venation of two dicotyledonous leaves and the arterial vasculature of the rat neocortex).

\subsection{Hierarchical decomposition of generated networks}

In this section we will consider various classes of architectural models. These computer generated networks were produced according to various predetermined rules. 
We use the hierarchical decomposition and associated metrics to quantify various aspects of the architecture, demonstrate what the metrics reveal about the graph organization and understand the effects of the finite size, boundaries and of noise.

The networks used in this section are presented in Fig.\ref{figure3}(a). The underlying geometry, link connectivity and point-wise link weight distribution are identical in every model. The architecture is solely defined by the building rule according to which the link weight values are assigned on the network. In the \textit{gradient} model in  Fig.\ref{figure3}, the link weights are distributed according to the link center Euclidean distance from the left-most vertex, creating a smooth gradient of link weight. In the \textit{peaks} model, the thick links are concentrated around seven equidistant peaks. In the \textit{nested} model, the straight lines defined by the underlying link connectivity are ordered based on a self similar subdivision scheme: the lines on the boundaries and center are assigned order $k=1$, the lines bisecting order $k=1$ lines are assigned order $k=2$ etc. 
The link edges are similarly ordered according to weight, and then distributed to the ordered straight lines so that higher thickness links occupy lower order lines. This produces a hierarchical self-similar pattern, characterized by long range order in the link weights.  The \textit{nested5} model demonstrates an instance of a class of models that is produced by the nested model, choosing five lines at random and randomly permuting their order. Similarly, the model \textit{nested10}, is derived from the nested model by swapping 10 lines at random. The \textit{random lines} model is produced by a random permutation of all the lines. Finally, the model \textit{random links} is produced by random assignment of the weights to the links and exhibits no log-range order.

\begin{figure*}[t!] 
\begin{center}
\includegraphics[width=6.7in]{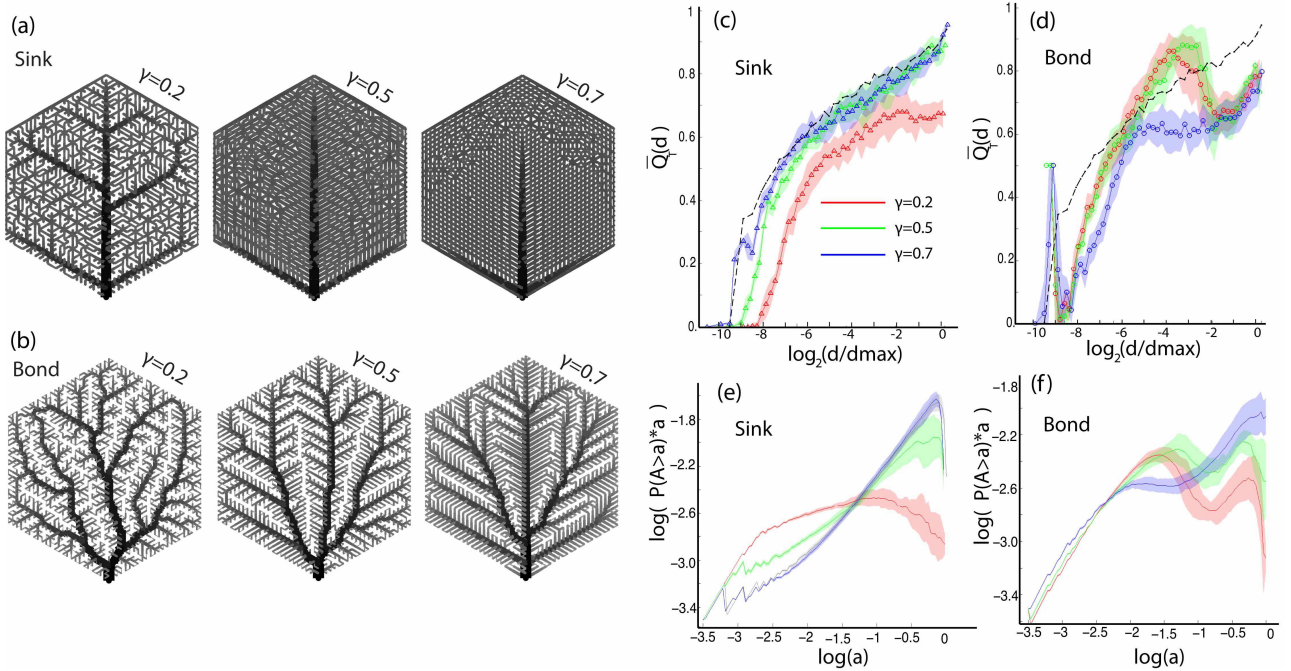}
\caption{\label{figure4}
\textbf{(a)} Optimized networks, fluctuations in the load (sink model). Instances of optimized graphs ($\gamma=0.2,0.5,0.7$) when the load is concentrated at a single, moving, point.
\textbf{(b)} Optimized networks, robustness to damage (bond model). Instances of optimized graphs ($\gamma=0.2,0.5,0.7$) when robustness is required under the presence of random damage. 
\textbf{(c)} Asymmetry of sink model. 
\textbf{(d)} Asymmetry of bond model. The average asymmetry $\bar{Q}_T(d)$ is plotted as a function of the normalized subtree degree $d/d_{\max}$. Red line: $\gamma=0.2$. Green line: $\gamma=0.5$. Blue line: $\gamma=0.7$. Black dashed line: random links model. The colored area represents the standard error after averaging over 20 realizations of each model. 
\textbf{(e)} Adjusted cumulative size distribution, sink models. The gray line overlayed on the blue, $\gamma=0.7$ line is the random links model.
\textbf{(f)} Adjusted cumulative size distribution, bond models.
 The adjusted cumulative size distribution $P(A>a)*a$ is plotted for $\gamma = 0.2,\; 0.5,\; 0.7$ (red, green, blue respectively) 
 The adjusted cumulative size distribution is averaged over 20 realizations for the bond, sink and random edges model. The colored area represents the standard error after averaging over 20 realizations of each model. 
} 
\end{center}
\end{figure*}

In Fig.~\ref{figure3}(b) we plot the Asymmetry $Q_T(t_j)$ for the architectural models termed nested (blue), gradient (magenta) and peaks (green). Ignoring the nesting tree vertices that correspond to the phantom boundary loops, the nesting tree of the gradient model is a purely additive tree of the type shown in Fig.~\ref{figure2}(e)(i${}_1$). The nesting tree of the nested model is similar to Fig.~\ref{figure2}(e)(ii${}_1$), however, the iterated building unit is composed of four sequentially joining elements, rather than just two joining nodes.

We have analytically calculated the Asymmetry of Eq.~\ref{Eq.asymmetry} for the infinite gradient and nested models (no boundaries). For the gradient model, it can be trivially found to be:
\begin{equation}
Q_T(t_j) = 1-\frac{2}{d(t_j)}.
\end{equation}
For the nested model, the analytical expression in closed form is more complicated and presented in the supplemental material.
To demonstrate the effects of the boundary in the Asymmetry of the nested and gradient model, we overlay the theoretical predictions on the finite size numerical results of Fig.~\ref{figure3}(b). For the gradient model, where a large number of low order loops break directly to the boundary, we notice a deviation of the actual measured finite size Asymmetry from the theoretical one. This deviation is mostly noticeable for small $d$. In the nested model case, there are no low level loops that join the exterior during the initial stages of the decomposition, so the finite size effects produce a deviation from the theoretical graph only at high degrees $d$. The damped fluctuations in the Asymmetry of the nested model are a signature of the model's self similarity. 
The asymptotic relaxation value of these fluctuations is indicative of the iterative building block of this architecture.

The Asymmetry plot of the peaks model follows closely the one of the gradient model, but, at approximately $d\simeq 2^{6.5}$ there is a marked change of monotonicity. This is the characteristic scale where the architecture of the model changes qualitatively. 
Until that scale, the architecture was predominately additive, with smaller loops sequentially joining larger ones, and the Asymmetry curve followed qualitatively that of the gradient model.
The Asymmetry decreases when the six separate, large size segments, represented in the inset graph with different colors, join. After those events take place during the hierarchical decomposition smaller loops with stronger edges continue to sequentially join creating a pattern in the Asymmetry plot that is again reminiscent of the gradient model.
The Asymmetry can be used to identify characteristic length scales in graphs where major changes in the architecture take place.


All the three models shown on Fig.~\ref{figure3}(b) are deterministic, with relatively simple architectures. Models such as the random links or the random lines model exhibit a much more complex asymmetry profile, as shown in Fig.~\ref{figure3}(c). 
The asymmetry values in that case are drawn from a distribution the properties of which reflect the architecture in question. Calculating mutual information and comparing density maps such as the ones shown in the inset of Fig.~\ref{figure3}(c) can provide a statistically meaningful way to examine the null hypothesis if two random graphs belong to the same architectural class. An extensive statistical comparison of the different architectural models is beyond the scope of this work. Alternatively, we calculate the average Asymmetry (\ref{Eq.average_asymmetry}), plotted in Fig.~\ref{figure3}(c) with the red and cyan solid lines and in Fig.~\ref{figure3}(d). The exact average asymmetry of each realization of the random models depends on the details of the noise. In Fig.~\ref{figure3}(d) we plot the mean $\bar{Q}_T(d)$ over 20 realizations of the nested (blue line), nested5 (orange), nested10 (light blue), random lines (red) and random links (cyan) models. The colored area represents the standard error.

The nested5 and nested10 models represent intermediate models between the nested and the random lines architecture, with progressively increasing disorder (and Asymmetry) as the number of lines that have been swapped becomes greater. The nested, nested5, nested10 and random lines model are architectures with long range order in the link strength, qualitatively significantly different than the random links model, in which the link strength is uncorrelated. This difference in reflected in the Asymmetry values of the random lines and random links models (Fig.~\ref{figure3}(d)).


The cumulative size distribution of the generated models is presented in Fig.~\ref{figure3}(e) and Fig.~\ref{figure3}(f${}_{1,2}$). In particular, in Fig.~\ref{figure3}(e) we plot the cumulative size distribution of the peaks (green), gradient (magenta), nested (blue) and random links model (cyan). As anticipated, the gradient model follows a straight line of slope 1/2 (a small deviation for small $a$ is due to boundary effects). Kinks and discontinuities in the slope, like the ones seen in the peaks model curve, are indicative of qualitative changes in the architecture. The random lines and nested model curves are significantly different from the gradient model. We can robustly test for scale invariance by defining the adjusted cumulative size distribution $a \cdot P(A>a)$. According to Eq.~\ref{Eq.self_similar}, for self similar graphs like the nested model we expect this quantity to fluctuate around a constant value, and the shape of the fluctuations are indicative of the iterative building block of the nested model.

 In Fig.~\ref{figure3}(f${}_1$) we plot $a \cdot P(A>a)$ for the random links model, and in Fig.~\ref{figure3}(f${}_2$) for the various nested and random lines models. As expected, the curves for all realizations of the self similar models fluctuate around a straight line. The periodicity of the curve can reveal the size of the architectural unit of the self similar network. The deviation from a straight line for large $a$ is due to boundary effects. As the disorder increases, the periodicity becomes less pronounced, and disappears at the random lines model.

\subsection{Hierarchical decomposition of optimized networks}

In this section we use the hierarchical decomposition method and the nesting tree to analyze the output of the optimization routines presented in \cite{Katifori:2010p2741}. Here, unlike the architectural models presented earlier, the building rules according to which the networks were constructed are not a-priori known. However, the functional purpose of the networks is known, as they are the (local) minima of global energy functions. The two models under consideration are a robustness to damage (broken bond) and fluctuations in the load (sink) model.

Modeled as electrical (or equivalently water distribution) grids, the networks transport load from the root (bottom center vertex in the networks of Fig.~\ref{figure4}(a)) to other nodes in the network. In the "bond" model, the root has to distribute the load evenly to all the vertices, even if a random single bond is removed (robustness to damage). In the fluctuating sink model, instead of a uniform distribution of sinks there is a single sink the position of which moves across the network. The cost to build the network is determined by a function $K=\sum C^{\gamma}$ and is set to a constant in each case. The parameter $\gamma$ quantifies the ``economy of scale'', i.e. how relatively expensive is a high conductivity edge compared to a smaller edge. The link thickness of the graphs shown in Fig.~\ref{figure4}(a) represents the bond conductivities, which are determined by optimizing for the total network power dissipation (results are shown for $\gamma=$ 0.2, 0.5 and 0.7 ).

\begin{figure*}[t!] 
\begin{center}
\includegraphics[width=6.8in]{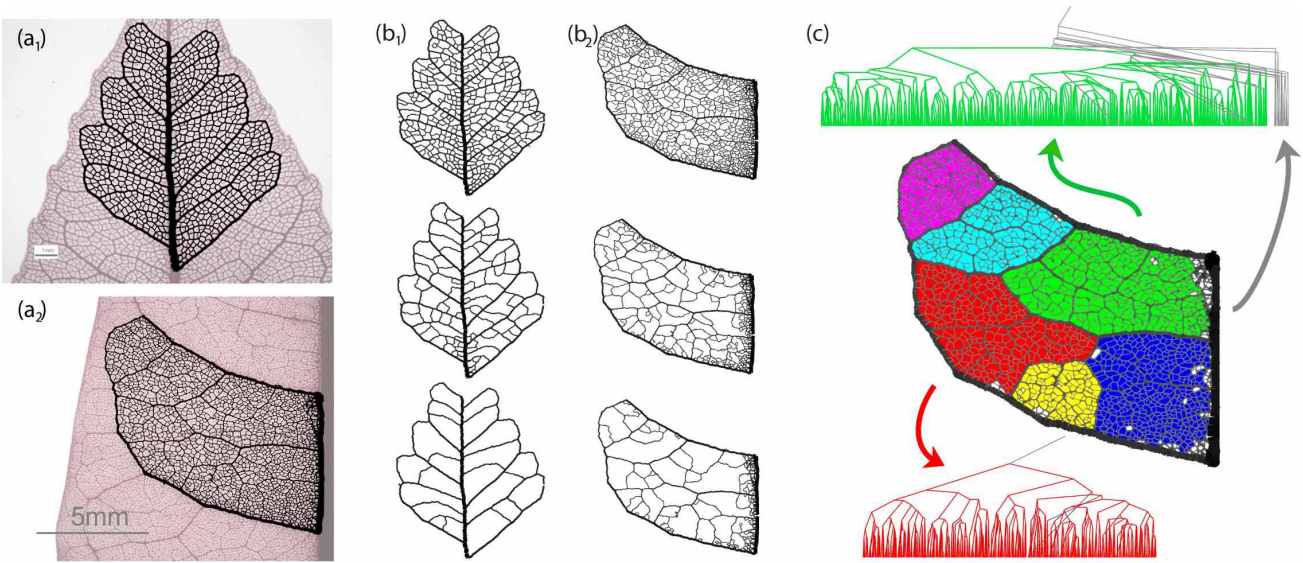}
\caption{\label{figure5}
\textbf{(a)} Segments of digitized leaf vasculature. The image of the skeletonized leaf has been overlayed with the digitized portion of interest. (a${}_1$) Bursera teconata, (a${}_2$) Protium heptaphyllum. Images courtesy of Douglas Daly, New York Botanical Gardens. 
\textbf{(b)} Hierarchical decomposition of Bursera and Protium. (a${}_1$) Bursera, (a${}_2$) Protium. Top to bottom remaining loop at three different, progressively higher thickness cutoffs. Notice the persistent minor loops at the proximity of the major veins.
\textbf{(c)} Segmentation of Protium heptaphyllum and associated tree representation. The protium intercostal area area has been separated to six color-coded sectors, as identified by hierarchical decomposition. The associated tree representation for that sector is shown for the green and red sector. The non-colored (white) areas of the graph and associated gray links on the tree representation correspond to high asymmetry nodes of the tree representation. Note how the high asymmetry areas are concentrated near major leaf veins.
} 
\end{center}
\end{figure*}


The Asymmetry plots demonstrate the strong statistical similarity of the sink $\gamma=0.5$ and $\gamma=0.7$ models with the random links model at intermediate and large scales (Fig.~\ref{figure4}(c)). For the bond models, the  $\gamma=0.2$ follows closely the $\gamma=0.5$ optimum, and they both exhibit a marked change in monotonicity at larger scales. The overall asymmetry increases with $\gamma$ in the sink model, whereas there appears to be a significant qualitative change in the architecture between the $\gamma=0.7$ and $\gamma=0.2, 0.5$ of the bond model.
Here it should be noted that the asymmetry metric, as defined here, does not depend on the actual numerical value of the bond strengths, just the absolute ordering on the lattice. The sink model network for $\gamma=0.5$, $0.7$ appears uniform as the smaller conductivity values are similar in value, however, architecturally the network is similar to the random model of Fig.~\ref{figure3}(a).


The adjusted cumulative size distribution shown in Fig.\ref{figure4}(e),(f), overall qualitatively reproduces the findings of the Asymmetry. The bond model for $\gamma=0.7$ exhibits a small size plateau. The sink $\gamma=0.7$ model follows a similar curve as the one of the random links model. Note the change of monotonicity in the bond $\gamma=0.2$ and $\gamma=0.5$ model. This indicates a change of architecture from primarily additive to primarily multiplicative nestedness. 

\begin{figure*}[b!] 
\begin{center}
\includegraphics[width=6.8in]{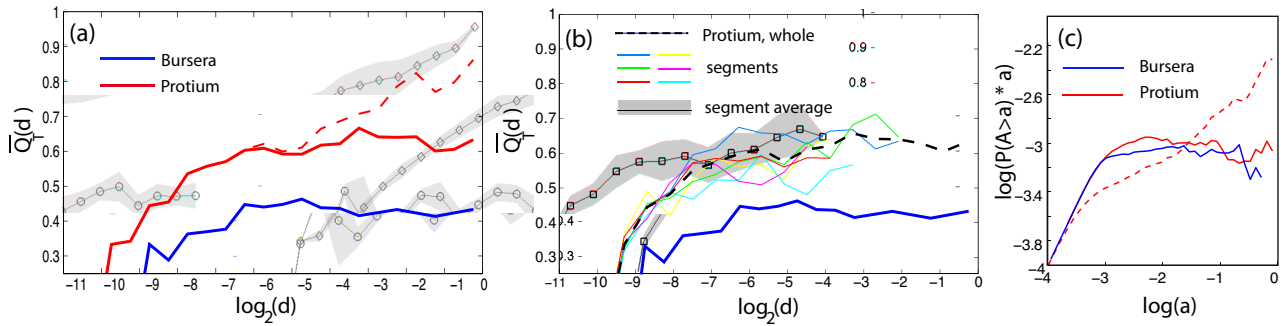}
\end{center}
\caption{\label{figure6}
\textbf{(a)} Asymmetry of Bursera and Protium intercostal areas. The average asymmetry $\bar{Q}_T(d)$ is plotted as a function of the normalized subtree degree $d$. Red solid line: Protium, cleaned. Red dashed line: Protium, full graph. Blue line: Bursera. Dark diamonds: random edges model.  Dark circles: nested model.
\textbf{(b)} Asymmetry of Protium intercostal segments. $\bar{Q}_T(d)$ is plotted as a function of the normalized subtree degree $d$. Black dashed line: Protium, cleaned. Red, blue, green, magenta, cyan, yellow lines: Protium segments, colorcoded as in Fig.~\ref{figure5}. Gray squares: average of segment asymmetry with standard error.
\textbf{(c)} Adjusted cumulative size distribution, Bursera and Protium. Red solid line: Protium, cleaned. Red dashed line: Protium, full graph. Blue line: Bursera. 
}
\end{figure*}


\subsection{Hierarchical decomposition of  natural networks}

In this section we apply the methodology we developed so far for two real examples, a leaf from Bursera teconata and a leaf from Protium heptaphyllum, show on Fig.\ref{figure5}. The leaves have been cleared and stained by the group of D. Daly in the New York Botanical Gardens, who provided us with high resolution images of the specimens. We reconstructed and digitized the vasculature of leaves using custom made software that we have developed to translate the pixel values information to a collection of nodes and edges on which we can perform hierarchical decomposition.

In Fig.~\ref{figure5} we show the reconstructed portion of the leaves, overlayed on the digital image from which it was acquired. A non-uniform staining or illumination of the specimen can introduce bias to the reconstruction algorithm and certain neighborhoods of the reconstructed graph might appear to have spuriously large weights. 
In particular, executing an initial decomposition step on the two networks of in Fig.~\ref{figure5}(b${}_1$) and (b${}_1$), we can easily see that unlike the Bursera, the Protium sample appears to have strong loops of smaller size concentrated around major veins. A careful inspection of the actual specimen is necessary to determine whether the origin of this bias is due to differential staining or this effect is of true biological origin. Although problems like this can be dealt before the digitization step in a number of ways (such as a variable threshold), here we will not discuss this, but will accept the input data at face value, and discuss cleaning methods that can be applied post the digitization stage.


A hierarchical decomposition of the intercostal area of Bursera allows us to identify high level nodes of the nesting tree that correspond to major loops.
We use the nesting tree to identify a natural segmentation of the graph to six major areas which we plot in Fig.~\ref{figure5}(c) along with the corresponding nesting subtrees for two of those sections. 
The histogram of the partition asymmetry $q$ defined on the nodes of the nesting tree has a local minimum at approximately $q\simeq 0.97$. This value can serve as a natural cutoff for data cleaning, In the nesting trees of Fig.~\ref{figure5}(c), we color the links of the subtree upstream of the nodes with partition asymmetry higher than $0.97$ with gray. The corresponding high asymmetry loops are colored white in the original graph.
We see that indeed the high symmetry loops are consistently concentrated around major veins.

The asymmetry curve $\bar{Q}_T$ of the intercostal area of Bursera, reaches a plateau. On the contrary, the Protium asymmetry does not relax to a constant value. However, if we clean the sample by disregarding the high asymmetry nodes with $q>0.97$, we see that Protium asymmetry curve similarly reaches a plateau, which is nevertheless higher than Bursera, indicating an architectural model based on more additive than multiplicative building blocks compared to Bursera.
We can calculate the asymmetry for each individual segment of Protium in Fig.~\ref{figure5} and see that, as expected, the different segments exhibit the same architecture and the asymmetry curves relax to a value of approximately $\bar{Q}_T(d\rightarrow 1) \simeq 0.6$, significantly different than the value of 0.45 of the Bursera. 



The cumulative size distributions of Fig.~\ref{figure6} qualitatively  follow the observations from the asymmetry plots. The cleaned Protium curve, as well as the Bursera curve, both reach a plateau, however the cumulative asymmetry cannot effectively distinguish between the two speciments.

\begin{figure}[b!] 
\begin{center}
\includegraphics[width=3.2in]{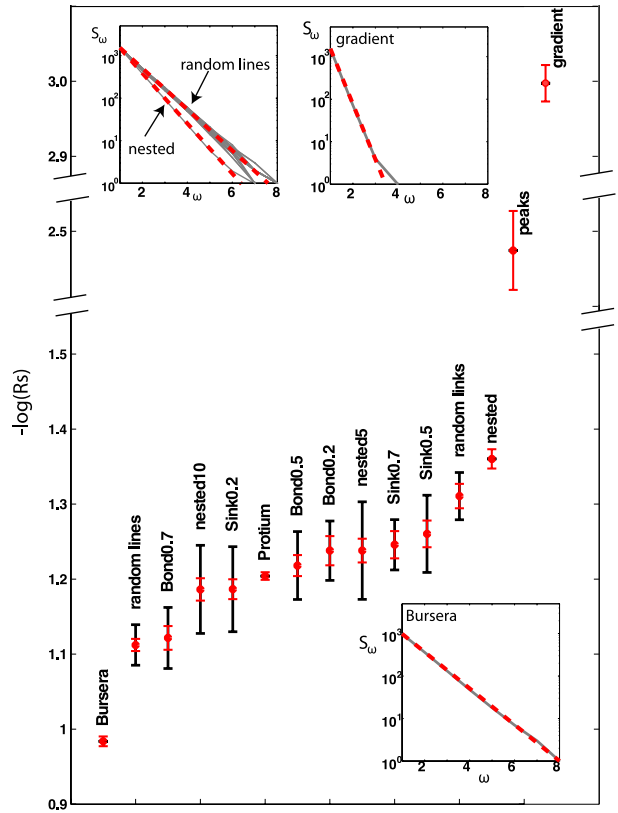}
\caption{\label{figure7}
Strahler bifurcation ratio for the various generated, optimized and natural graphs presented in this paper. Red error bar: standard error of the linear regression fit (represents goodness of linear fit). Black error bar: standard deviation of the  logarithm of the bifurcation ratio (average over 20 realizations). Insets: Number of Strahler streams $S_{\omega}$ of order $\omega$ as a function of $\omega$ for the random lines, nested and gradient model and the Bursera leaf. Note that in each case, the $S_{\omega}$ follows closely an inverse geometric progression with $\omega$ (shown with the red dashed line).
} 
\end{center}
\end{figure}

\subsection{Strahler bifurcation ratio}

The Strahler bifurcation ratio (\ref{Eq.Strahler_bifurcation}) defined on the nesting tree can provide a metric to quantify the overall nestedness of graphs. Since the Strahler law of stream numbers is an inevitable reality for most trees, it is possible to fit the plots $\log(S_{\omega})$ versus $\log \omega$ with a straight line the slope of which will determine the logarithm of the Strahler bifurcation ratio $R_s$ for the whole graph. Examples of this fit are shown in the inset of Fig.~\ref{figure7}. The best fit is found in the least squares sense, and it is forced to pass through $(1,S_1)$ ($S_1$ is equal to the total number of ultimate loops, or leaf nodes in the nesting tree). The data point for $\max(\omega)$ is discarded, as it is very sensitive to noise.

In Fig.~\ref{figure7} we plot the Strahler bifurcation ratios for all the graphs presented in this paper. For the architecture or the optimization models that are not deterministic each realization of the graph will produce a different $R_s$. In those cases, we plot  $\langle R_s \rangle$, the average bifurcation ratio over 20 realizations, with the black error bar being the standard deviation. The red error bar represent the (average) goodness of the linear fit. Notice the extent of the red error bar for the gradient and peaks models. 

The Strahler bifurcation ratio can clearly distinguish between the strongly multiplicatively nested Bursera and additively nested gradient model, but, with our current implementation it could not sufficiently distinguish between many of the models presented in this work. A major drawback of $R_s$ is that it is a single number which is inherently unsufficient to capture the complexity of networks whose architectural properties do not necessarily remain the same over all lengthscales.

\subsection{The rat brain}

The analysis and methodology presented in this work can be useful not only for leaves, but any other, biological or man made, planar graphs. A notable example is the arterial vasculature of the rodent neocortex which forms a planar network with multiple loops \cite{Blinder:2010p2746}. We extracted the diameters of the arterial blood vessels from a composite rat brain image provided to us by the Kleinfeld group in UCSD and augmented the connectivity information in \cite{Blinder:2010p2746} to obtain a weighted map of the arterial vasculature of the rat brain. Although the resolution of the image in our disposal does not allow us to determine the vein widths with absolute confidence, we were able to identify major vascular sectors and determine that, according to the data at hand, the architecture of the network in question is primarily additive than multiplicative.

\section{Discussion}

We have presented a framework that allows us to quantify the hierarchical organization of predominately loopy architectures. Our {\em hierarchical decomposition} consists of three iteratively repeated steps:
\begin{enumerate}
\item pruning of the tree-like components
\item ordering of the edges
\item removal of the thinnest edge
\end{enumerate}
This framework relies on the mapping of loopy planar graphs and their hierarchical decomposition to binary nesting trees. The nesting tree is subsequently used to quantify the architectural organization of the original graph. A number of quantities that reflect various aspects of the graph organization can be defined on the nesting tree, each with each own advantages and disadvantages. In this work we presented results for three such quantities, the Asymmetry $\bar{Q}_T$, the cumulative size distribution and the Strahler bifurcation ratio. The Asymmetry is a bottom-up approach that assigns a number to every composite loop at each scale. This number is a weighted average of the nestedness of the architecture of the portion of the graph enclosed in that loop. Depending on the averaging window, two different architectures of a high degree loop can map to the same $\bar{Q}_T$ value. On the contrary, the cumulative size distribution performs better in differentiating architectures at the high levels of organization. The larger number of low level loops frequently results in washing out interesting features of the structures at smaller scales.

\begin{figure}[b!] 
\begin{center}
\includegraphics[width=3.2in]{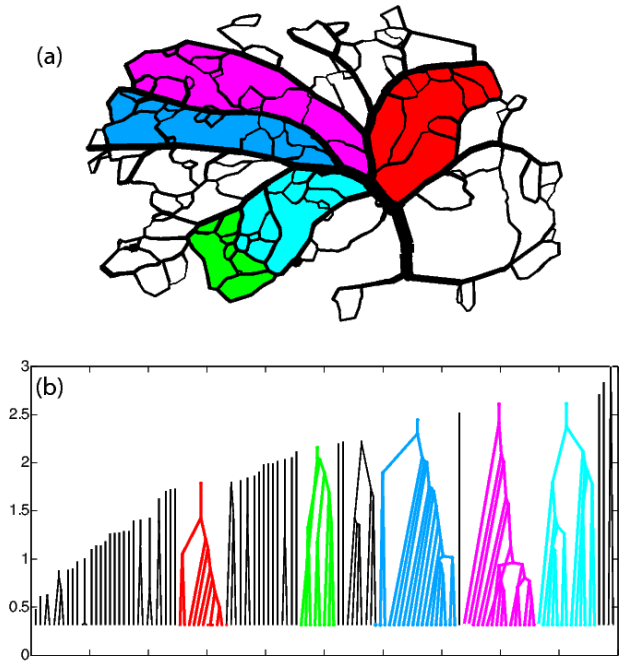}
\caption{\label{figure8}
Digitized arterial vasculature of rat neocortex and corresponding nesting tree representation. (a)The arterial network forms a planar graph. Different segments of the network, as identified by hierarchical decomposition are represented by different colors. (b) Nesting tree of the digitized network. the highlighted segments of the network are color-coded.
} 
\end{center}
\end{figure}

These observations are demonstrated in the sink and bond model Asymmetries and cumulative distributions of Fig.~\ref{figure4}.  For example, the Asymmetry of the sink $\gamma=0.7$ and bond models (Fig.~\ref{figure4}(c), (d)) has a local maximum, a feature that is absent from the adjusted cumulative size distribution (Fig.~\ref{figure4}(e), (f)). Similarly, the Asymmetry of all bond models is indistinguishable for large scales, whereas the cumulative size distribution can statistically distinguish these models.

Depending on the weight function $w_j$, the Asymmetry can be used to define a single number that encompasses information about the whole architectural organization (e.g. by calculating $\bar{Q}_T(d\rightarrow d_{\max})$) rather than to examine the architecture at all levels of organization. Such a number would be meaningful only for graphs with some degree of self-similarity.
The Strahler bifurcation ratio $R_s$ is is used to describe the overall architecture, but it does not perform well for complex architectures. We have examined the Strahler bifurcation ratio as a function of the Strahler order $\omega$ and degree $d$ as an attempt to extract information about the scale dependent organization of the graph.We have found that the result is very sensitive to noise, especially at high $\omega$. 

The metrics presented in this paper focus on the metric topology of the structure but they do not explicitly capture any information about the geometry of the network. The cumulative area distribution depends on the area of the terminal loops (the areoles of a leaf vein network). The cumulative area distribution follows closely the cumulative degree distribution provided that the terminal loops are not substantially polydisperse. It is evident we can supplement the descriptions presented here with more detailed geometrical analysis, in which some aspects of the geometry of the closed loops is kept, such as e.g. an approximating SVD ellipsoid, which can be used to define a major axis and an excentricity. We can then incorporate such geometrical information into the analysis of nesting, i.e., relationships defining what is the average orientation of subloops in relation to the parent loops. Such detailed geometrical analysis, however, will evidently be subordinate to the coarser topological analysis we have presented here. 

A big part of our extensive understanding of fluvial networks is due to the development of methodology to characterize and quantify tree architectures. Accordingly, progress in understanding loopy networks, which are ubiquitous in both natural and man made structures, is contingent on our ability to measure their hierarchical architecture. 
The approach presented in this work provides a robust mathematical description of the network architecture, applicable to leaf venation and other loopy distribution (and structural) structures. It can be used to characterize the {\sl in silico} networks obtained from computer simulations as well as to perform quantitative statistical comparisons between theory and experiment. As such, it can provide an invaluable tool in deciphering the functional signficance of the loopy networks and possibly their developmental origin.

\section*{Acknowledgments}
This work was supported in part by the National Science Foundation under Grant PHY-1058899. 
EK would like to acknowledge support from the Burroughs-Welcome Career at the Scientific Interface Award and the Raymond and Beverly Sackler Fellowship. The authors would like to thank D. Daly from the New York Botanical Garden for kindly providing us with cleared leaf images, and P. Blinder and the Kleinfeld group at University of California San Diego, for making available the rat neocortex vasculature data to us. 

During preparation of this manuscript the authors became aware of the work of Mileyko at al., concurrently submitted for publication. 

\bibliography{Katifori_Architecture_preprint}

\begin{thebibliography}{10}

\bibitem{Ellis2009}
Ellis B, {et~al.}
\newblock (2009) \emph{Manual of Leaf Architecture}
\newblock (Cornell University Press,Ithaca, NY).

\bibitem{Tero2010}
Tero A, {et~al.}
\newblock (2010) Rules for biologically inspired adaptive network design.
\newblock \emph{Science} 327:439--442.

\bibitem{Blinder:2010p2746}
Blinder P, Shih AY, Rafie C, Kleinfeld D
\newblock (2010) Topological basis for the robust distribution of blood to
  rodent neocortex.
\newblock \emph{Proc Natl Acad Sci USA} 107:12670--12675.

\bibitem{Melville1969}
Melville R
\newblock (1969) Leaf venation patterns and the origin of the angiosperms.
\newblock \emph{Nature} 224:121--125.

\bibitem{Katifori:2010p2741}
Katifori E, Szollosi GJ, Magnasco MO
\newblock (2010) Damage and fluctuations induce loops in optimal transport
  networks.
\newblock \emph{Phys Rev Lett} 104:048704.

\bibitem{Corson:2010p2818}
Corson F
\newblock (2010) Fluctuations and redundancy in optimal transport networks.
\newblock \emph{Phys Rev Lett} 104:048703.

\bibitem{Roth-Nebelsick2001a}
Roth-Nebelsick A, Uhl D, Mosbrugger V, Kerp H
\newblock (2001) Evolution and function of leaf venation architecture: A
  review.
\newblock \emph{Ann Bot} 87:553--566.

\bibitem{Couder2002}
Couder Y, Pauchard L, Allain C, Adda-Bedia M, Douady S
\newblock (2002) The leaf venation as formed in a tensorial field.
\newblock \emph{Eur Phys J B} 28:135--138.

\bibitem{Dimitrov2006}
Dimitrov P, Zucker SW
\newblock (2006) A constant production hypothesis guides leaf venation
  patterning.
\newblock \emph{Proc Natl Acad Sci USA} 103:9363--9368.

\bibitem{Fujita2006}
Fujita H, Mochizuki A
\newblock (2006) The origin of the diversity of leaf venation pattern.
\newblock \emph{Dev Dynam} 235:2710--2721.

\bibitem{Rinaldo2006}
Rinaldo A, Banavar JR, Maritan A
\newblock (2006) Trees, networks, and hydrology.
\newblock \emph{Water Resour Res} 42:W06D07.

\bibitem{Corson:2009p3346}
Corson F, Adda-Bedia M, Boudaoud A
\newblock (2009) In silico leaf venation networks: Growth and reorganization
  driven by mechanical forces.
\newblock \emph{J Theor Biol} 259:440--448.

\bibitem{HORTON:1945p3487}
Horton R
\newblock (1945) Erosional development of streams and their drainage basins -
  hydrophysical approach to quantitative morphology.
\newblock \emph{Geol Soc Am Bull} 56:275--370.

\bibitem{STRAHLER:1952p3188}
Strahler A
\newblock (1952) Hypsometric (area-altitude) analysis of erosional topography.
\newblock \emph{Geol Soc Am Bull} 63:1117--1141.

\bibitem{VANPELT:1992p2856}
VanPelt J, Uylings HBM, Verwer RWH, Pentney RJ, Woldenberg MJ
\newblock (1992) Tree asymmetry - a sensitive and practical measure for binary
  topological trees.
\newblock \emph{B Math Biol} 54:759--784.

\bibitem{Rinaldo:1998p259}
Rinaldo A, Rodriguez-Iturbe I, Rigon R
\newblock (1998) Channel networks.
\newblock \emph{Annu Rev Earth Pl Sc} 26:289--327.

\bibitem{Albert2002}
Albert R, Barabasi A
\newblock (2002) Statistical mechanics of complex networks.
\newblock \emph{Rev Mod Phys} 74:47--97.

\bibitem{Newman2003}
Newman MEJ
\newblock (2003) The structure and function of complex networks.
\newblock \emph{SIAM Review} 45:167--256.

\bibitem{Barrat2004}
Barrat A, Barthelemy M, Pastor-Satorras R, Vespignani A
\newblock (2004) The architecture of complex weighted networks.
\newblock \emph{Proc Natl Acad Sci USA} 101:3747--3752.

\bibitem{Boccaletti:2006p256}
Boccaletti S, Latora V, Moreno Y, Chavez M, Hwang DU
\newblock (2006) Complex networks: Structure and dynamics.
\newblock \emph{Phys Rep} 424:175--308.

\bibitem{Costa2007}
da~F~Costa L, Rodrigues FA, Travieso G, Boas PRV
\newblock (2007) Characterization of complex networks: A survey of
  measurements.
\newblock \emph{Adv Phys} 56:167--242.

\bibitem{Shao:2009p3561}
Shao J, Buldyrev SV, Braunstein LA, Havlin S, Stanley HE
\newblock (2009) Structure of shells in complex networks.
\newblock \emph{Phys Rev E} 80:036105.

\bibitem{Sales-Pardo2007}
Sales-Pardo M, Guimera R, Moreira AA, Amaral LAN
\newblock (2007) Extracting the hierarchical organization of complex systems.
\newblock \emph{Proc Natl Acad Sci USA} 104:15224--15229.

\bibitem{Ang2005}
Ang WK, Jowitt P
\newblock (2005) Some new insights on informational entropy for water
  distribution networks.
\newblock \emph{Eng Optimiz} 37:277--289.

\bibitem{Rolland-Lagan2009}
Rolland-Lagan AG, Amin M, Pakulska M
\newblock (2009) Quantifying leaf venation patterns: two-dimensional maps.
\newblock \emph{Plant J} 57:195--205.

\bibitem{Blonder:2011p3510}
Blonder B, Violle C, Bentley LP, Enquist BJ
\newblock (2011) Venation networks and the origin of the leaf economics
  spectrum.
\newblock \emph{Ecol Lett} 14:91--100.

\bibitem{TAKAYASU:1988p3366}
Takayasu H, Nishikawa I, Tasaki H
\newblock (1988) Power-law mass-distribution of aggregation systems with
  injection.
\newblock \emph{Phys Rev A} 37:3110--3117.

\bibitem{Paik:2007p946}
Paik K, Kumar P
\newblock (2007) Inevitable self-similar topology of binary trees and their
  diverse hierarchical density.
\newblock \emph{Eur Phys J B} 60:247--258.

\bibitem{KIRCHNER:1993p265}
Kirchner J
\newblock (1993) Statistical inevitability of horton laws and the apparent
  randomness of stream channel networks.
\newblock \emph{Geology} 21:591--594.

\end{thebibliography}




\end{document}